\documentclass[aps,preprint,superscriptaddress,showpacs,amsmath,amssymb,endfloats* ]{revtex4}

\usepackage{graphicx}
\usepackage{bm}

\begin{document}

\title{Analysis of a free oscillation atom interferometer}
\author{Rudra P. Kafle}
\affiliation{Department of Physics, Worcester Polytechnic Institute,
100 Institute Road, Worcester, Massachusetts 01609, USA}
\author{Dana Z. Anderson}
\affiliation{Department of Physics and JILA, University of Colorado
and National Institute of Standards and Technology, Boulder,
Colorado 80309-0440, USA}
\author{Alex A. Zozulya}\email[]{zozulya@wpi.edu}
\affiliation{Department of Physics, Worcester Polytechnic Institute,
100 Institute Road, Worcester, Massachusetts 01609, USA}

\begin{abstract}
We analyze a Bose-Einstein condensate (BEC) - based free oscillation atom Michelson interferometer in a weakly
confining harmonic magnetic trap. A BEC at the center of the trap is split into two harmonics by a laser standing wave. The harmonics move in opposite directions with equal speeds and turn back under the influence of the trapping potential at their classical turning points. The harmonics are allowed to pass through each other and a recombination pulse is applied when they overlap at the end of a cycle after they return for the second time. We derive an expression for the contrast of the interferometric fringes and obtain the fundamental limit of performance of the interferometer in the parameter space.
\end{abstract}

\pacs{03.75.Dg, 37.25.+k}

\maketitle

\section{Introduction}\label{sec:intro}

Atom interferometers using cold atoms or Bose-Einstein condensates (BECs) can have very high sensitivities in comparison to their optical counterparts \cite{wang05}, and can find potential applications in field-sensing and precision measurements \cite{berman97}.  Atom interferometers can be more versatile than the optical ones and have been used to measure acceleration \cite{kasevich91}, rotations \cite{lenef97}, and dynamic polarizability of atoms \cite{deissler08}.

The first atom interferometry experiments with supersonic atomic beams were reported in Ref. \cite{carnal91, keith91}. The laser cooling techniques of neutral atoms developed in the 1980s \cite{phillips98} opened up the applications of cold atoms in atom interferometry.  Atom interferometry with cold atoms by projecting them in a vertical direction was used to measure acceleration due to gravity \cite{kasevich91}.

After the experimental realizations of BECs in dilute atomic gases in the mid-nineties \cite{anderson95, davis95, bradley95}, the horizon of atom interferometry has broadened. The atoms in BECs have a very narrow momentum distribution and hence can be controlled and manipulated more easily than the thermal atoms by using light waves. Moreover, all atoms in BECs are in the same quantum state and hence BECs are excellent coherent sources of matter waves. The interference of two independent condensates was first reported in Ref. \cite{andrews97}, in which two separate condensates were prepared in a double-well potential and allowed to interfere by switching off the potential and letting the condensates expand. Shin et al. \cite{shin04} showed trapped atom interferometry with a condensate prepared in an optical single-well potential and then coherently split into two by deforming the single-well into a double-well potential. This, as well as several other experiments \cite{wang05, schumm05, burke08, horikoshi07} on BEC-based atom interferometry, shows that condensates are good candidates for interferometric applications. BEC-based atom interferometers in Michelson geometry \cite{wang05,garcia06} and in Mach-Zehnder geometry \cite{torii00, horikoshi06} were realized recently.

The basic steps of guided atom interferometry are the following \cite{cronin09}: an atomic wave packet is split into two in a trap or a wave guide, the split wave packets are sent down two different paths, and recombined at the end of the interferometric cycle. For example, in a single-reflection atom Michelson interferometer,  a BEC in a zero momentum state $\psi_0$ is split at time $\tau = 0$ by a laser standing wave into two harmonics $\psi_+$ and $\psi_-$ \cite{wu05,wang05,stickney07}. The atoms in the $\psi_+$ harmonic absorb a photon from a laser beam with the momentum $\hbar k_l$ and re-emit into the beam with the momentum $-\hbar k_l$ (with $k_l$ being the wave number of the laser,)  thus acquiring velocity $ v_0 = {2\hbar k_l}/M$, where $M$ is the atomic mass. Similarly ,  an atom in the $\psi_-$ harmonic acquires velocity $ -v_0 = -{2\hbar k_l}/M$. At time $\tau = T/2$, where $T$ is the interferometric cycle time, a reflection pulse is applied to reverse the momenta of the harmonics. At time $\tau = T$, the two harmonics are subject to the action of a recombination pulse. After recombination, in general, the atoms populate all three harmonics $\psi_{0}$ and $\psi_{\pm}$. The number of atoms in each harmonic depends on the relative phase acquired during the interferometric cycle and can be used to deduce this phase.

Loss of contrast in a single reflection interferometer is primarily due to a coordinate-dependent phase acquired during the cycle.  This phase is caused by the confining potential and the velocity mismatch due to reflection pulses. To overcome this drawback, a double pass interferometer with reflection pulses was proposed and implemented in Ref.~\cite{garcia06}.  In this interferometer, each cloud travels in both arms of the interferometer before the clouds are finally recombined.  Because of the symmetry in the paths followed by the two clouds, the coordinate dependent phase is partially canceled.  But the reflection pulses still introduce a velocity mismatch that limits the performance of the interferometer \cite{stickney07}. The performance can further be improved in the geometry of a free oscillation interferometer \cite{segal10,burke08} that does not use the reflection pulses at all.  In this geometry, the split clouds turn around in the confining potential at their classical turning points and pass through each other twice before they are finally recombined.

Both double reflection and free oscillation interferometers are not suitable for measuring static environment like gravitational acceleration because both clouds accumulate equal environment-introduced phases during an interferometric cycle and hence the relative phase shift is zero. But they can be used for measuring environmental effects that can be controlled in time. For example,  Deissler et al. in Ref.  \cite{deissler08} measure the dynamic polarizability of $^{87}$Rb atoms with a double pass interferometer. Burke et al. in Ref. \cite{burke09} show that a double pass interferometer can be used as a  Sagnac interferometer to measure rotation using the Sagnac effect.

\begin{figure}
\includegraphics[width=8.6cm]{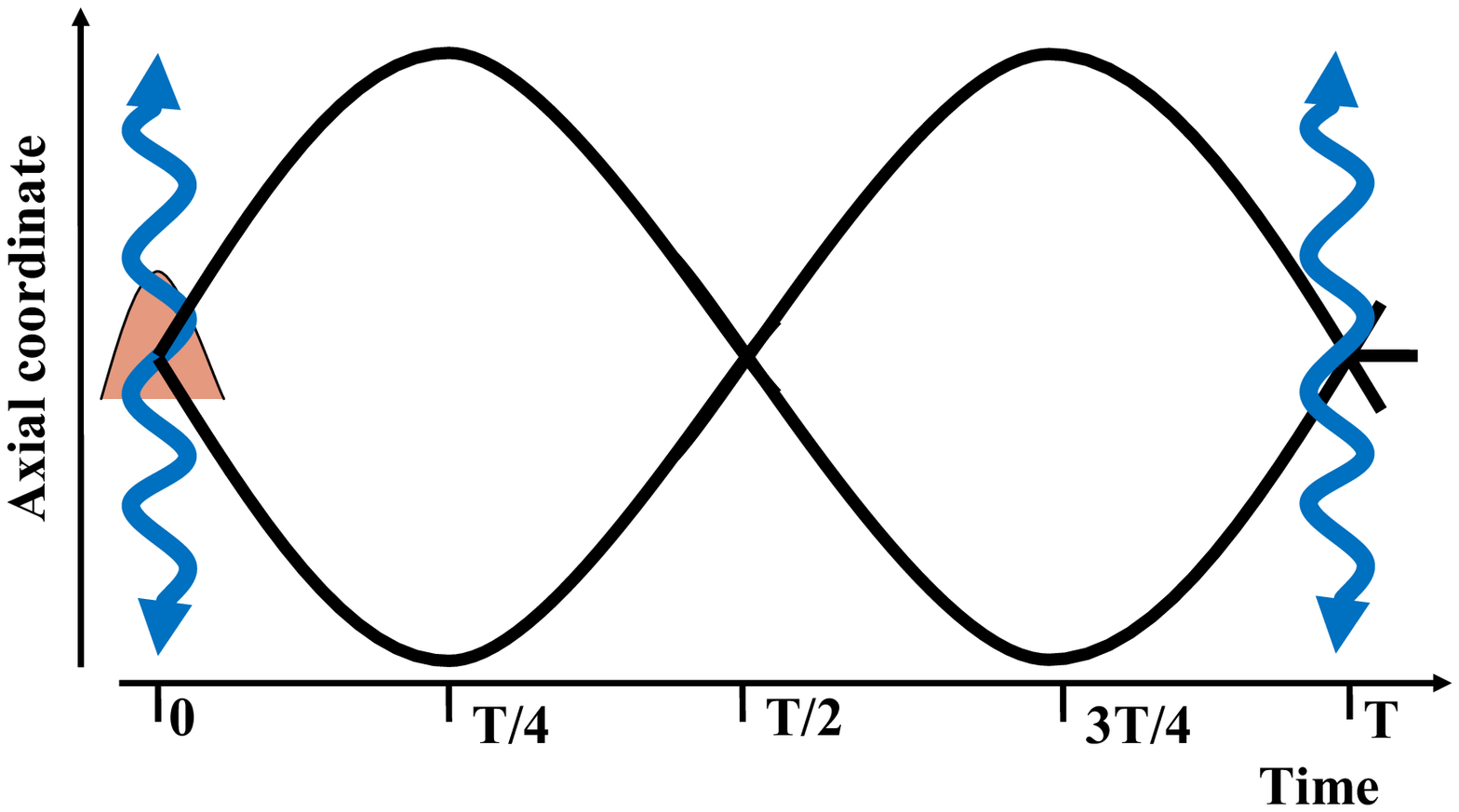}%
\caption{(Color online) \label{fig:Schematic of interferometer} A schematic of a free oscillation interferometer. \label{cartoon}}
\end{figure}

In this paper, we theoretically analyze a free oscillation atom Michelson interferometer in the framework of a mean field approach and obtain its limit of performance in the parameter space. This interferometer is schematically shown in Fig.~\ref{fig:Schematic of interferometer}. The solid lines are the paths followed by the two harmonics during an interferometric cycle and the vertical wavy lines represent the splitting and recombination laser pulses. The split condensates move in a weakly-confining harmonic trap and are reflected from their classical turning points. They pass through each other, reach the maximum excursions in the opposite arms and return again. The harmonics are recombined when they again overlap at the center of the trap. This interferometer has been experimentally realized in Ref. \cite{segal10,burke08} and  in a different (Mach-Zehnder) geometry in Ref. \cite{horikoshi07}. Horikoshi et al. \cite{horikoshi07} have shown that dephasing in this type of interferometer is suppressed. Nevertheless, there is still a fundamental limit on the performance of this geometry that is caused by the confining potential and the nonlinearity of the condensate, which is the subject of the present paper.

The rest of the paper is organized as follows: Section~\ref{sec:analytical_model} formulates the analytical model used to describe the interferometer. The equations of motion for the split condensates are derived and analyzed in Section \ref{sec:dynamics}.  The limits of performance of the free oscillation interferometer are discussed in Section \ref{sec:limit}.  The free oscillation interferometer is compared with the single and double reflection interferometers in Section \ref{sec:single-double}. Finally, conclusions are presented in Section \ref{sec:conclusions}.

\section{Analytical Model}\label{sec:analytical_model}

The results of this paper are obtained in the framework of a one-dimensional (1D) mean-field theory in Thomas-Fermi limit \cite{dalfovo99}.
A 1D model is a good approximation to the experimental situation \cite{wang05, horikoshi06, garcia06, horikoshi07, burke08, segal10}, where the BEC clouds are cigar-shaped with the largest dimension along the weak guiding direction of the trap and are moving along the same direction.

Specifically, we describe the evolution of a BEC in a weakly-confining parabolic potential of longitudinal frequency $\omega$ by the following
dimensionless Gross-Pitaevskii equation (GPE):
\begin{equation}\label{GPE}
i\frac{\partial}{\partial \tau} \psi(X,\tau) = \left[-\frac{\epsilon}{2} \frac{\partial^2}{\partial X^2}+ \frac{1}{2\epsilon}X^{2} + g_{1D}|\psi(X,\tau)|^2\right]\psi(X,\tau).
\end{equation}
 The axial coordinate $x$ is normalized to the initial longitudinal radius  $L_0$ of the condensate: $X = x/L_0$. The dimensionless time $\tau$ is given by the relation $\tau = \omega t$, where $\omega$ is the longitudinal frequency of the weakly-confining potential.
The strength of interatomic interactions is described by the parameter $g_{1D} = 2\omega_\perp a_s N/(\omega L_0)$ where $a_s$ is the $s$-wave scattering length, $N$ is the total number of atoms in the condensate and $\omega_\perp$ is the trapping angular frequency in the tightly confined transverse dimensions. Finally, $\epsilon = (a_0/L_0)^2$, where $a_0 = \sqrt{\hbar/(M\omega)}$ is the oscillator length along the longitudinal dimension. The wave function $\psi$ has been normalized to $1$. For details of derivation of Eq.~(\ref{GPE}) and its limits of applicability, see \cite{stickney07}.

The initial equilibrium size of the condensate in the Thomas-Fermi approximation \cite{dalfovo99} is given by :
\begin{equation}\label{size}
L_0 = \left(\frac{3\hbar \omega_\perp a_s N}{ M \omega^2} \right)^{1/3}.
\end{equation}
The wave function $\psi$ of the condensate after the splitting pulse is a superposition of two harmonics $\psi_{+}$ and $\psi_{-}$:
\begin{equation}\label{split_wavefn}
    \psi   = \frac{1}{\sqrt{2}}(\psi_+ + \psi_-).
\end{equation}
The wave functions $\psi_\pm$ have been normalized to $1$.

The densities $n_{\pm}$ and the phases $\phi_{\pm}$ of the harmonics $\psi_+$ and $\psi_-$ defined by the relations $\psi_\pm = \sqrt{n_\pm}\exp(i\phi_\pm)$, are represented as :
\begin{eqnarray} \label{density_phase}
n_\pm &=& \frac{3}{4R} \left[1-\left(\frac{X \mp X_0}{R}\right)^2\right],    \nonumber \\
\phi_\pm &=& (\phi_0)_\pm + \frac{1}{\epsilon} \left[\pm V(X\mp X_0) + \frac{G}{2}\left(\frac{X \mp X_0}{R}\right)^2\pm \frac{S}{6} \left(\frac{X \mp X_0}{R}\right)^3\right].
\end{eqnarray}

In Eq.~(\ref{density_phase}), $\pm X_{0}(\tau)$ are the positions of the centers of the two harmonics and $R(\tau) = L/L_0$ is their dimensionless radius. Since the splitting pulses act for a very short period of time, the harmonics' shape and
position immediately after  splitting are equal to those of the initial BEC at rest, i.e., $X_{0}(0) = 0$ and $R(0) = 1$.
In the expression for the total phase of the split BECs (the second equation of the Eq.~(\ref{density_phase})), the term $(\phi_0)_\pm$ is the phase accrued by the harmonics from the environment. The term $\pm V(X\mp X_0)$ is due to the motion of the two harmonics. The parameter $V(\tau)$ is the normalized speed $v$ of the harmonics, i.e.,  $V =  v/(\omega L_{0})$ with the initial value $V(0) = V_{0} = v_{0}/(\omega L_{0})$. The quadratic term proportional to $G$ appears because of dispersion of the harmonics. The cubic term proportional to $S$ is due to atom-atom interactions in the condensate. The quadratic and cubic phases are initially zero, $G(0) = S(0) = 0$ and evolve with time when the harmonics start propagating. The parabolic form of the density of the BEC clouds $n_{\pm}$ in Eq.~(\ref{density_phase}) implies the Thomas-Fermi limit, when the second derivative of the densities $n_{\pm}$ in Eq.~(\ref{GPE}) is neglected.

The Gross-Pitaevskii equation (\ref{GPE}), with the form of the density and phase of the two BEC clouds given by Eq. (\ref{density_phase}), has been previously investigated under various approximations both analytically and numerically in \cite{olshanii05,stickney07,burke08,stickney08}. In particular, the authors of Refs.~\cite{stickney07,stickney08} derived the set of coupled ordinary differential equations for the parameters $R(\tau)$, $X_{0}(\tau)$, $V(\tau)$, $G(\tau)$, $S(\tau)$, entering the expressions for the density and the phase of the BEC clouds given by Eq.~(\ref{density_phase}) (notations of Refs. \cite{stickney07, stickney08} are slightly different from those of the present paper). Validity of the analytical model has been confirmed by comparing solutions of these equations to the results following from direct numerical solution of the Gross-Pitaevskii equation (\ref{GPE}). The derivation of equations in Refs. \cite{stickney07, stickney08} has been based on representing the density and phase of the two BEC clouds (\ref{density_phase}) in terms of a truncated set of Legendre polynomials. In this paper we derive the set of equations analogous to that of Refs.~\cite{stickney07, stickney08} by analyzing equations of motion for the quantum-mechanical expectation values associated with the parameters $R(\tau)$, $X_{0}(\tau)$, $V(\tau)$, $G(\tau)$, and $S(\tau)$. This approach streamlines the derivation and allows for a greater insight into the physics. Additionally, it has the energy conservation law explicitly built in the formalism, greatly assisting further analysis. Finally, the new approach is more readily generalizable to two- or three-dimensional case. The set of derived equations is used to analyze the geometry of the free oscillation atom interferometer.

%
\section{Dynamics of the split condensates}\label{sec:dynamics}

\subsection{Equations of motion}
%
The time evolution of the expectation value $\langle \hat{A} \rangle = \langle \psi| \hat{A} | \psi \rangle = \int \psi^* \hat{A} \psi dx $  of a quantum mechanical operator $\hat{A}$ is governed by the equation
\begin{eqnarray}\label{Expectation_Equation}
\frac{d}{d\tau} \langle \hat{A} \rangle = i \langle \psi | [\hat{H}, \hat{A}]|\psi \rangle,
\end{eqnarray}
where $\hat{H}$ is the Hamiltonian of the system and $[\hat{H},\hat{A}] = \hat{H}\hat{A} - \hat{A}\hat{H} $ is the commutator. The Hamiltonians of the two BEC clouds after the splitting are given by the relation
\begin{eqnarray} \label{Hamiltonian}
H_{\pm} = \frac{\epsilon}{2} P^2 + \frac{1}{2\epsilon} X^2 + \frac{1}{2}g_{1D} (n_\pm + 2n_\mp).
\end{eqnarray}
The expectation values of the coordinate $X$, i.e., $\langle \psi | X | \psi \rangle$  and momentum $P$, i.e., $\langle \psi | P| \psi \rangle$, evaluated with respect
to the wave function $\psi_{+}$, are :
\begin{eqnarray} \label{Expect_values_XP}
 &&\langle X \rangle = X_0 ,\nonumber \\
 &&\langle P \rangle = V + \frac{S}{10R}.
\end{eqnarray}
Similarly, the expectation values of $X^2$, $P^2$, $n_+$ and $n_-$ are :
 \begin{eqnarray} \label{Expect_values_other}
&&\langle X^2 \rangle  =  X_0^2 + \frac{R^2}{5}, \nonumber \\
&&\langle P^2 \rangle  =  \left(V + \frac{S}{10R}\right)^2 + \frac{G^2}{5R^2} + \frac{2S^2}{175 R^2}, \nonumber \\
&&\langle n_+ \rangle  =  \frac{3}{5R},\\
&&\langle n_- \rangle  =  \frac{3}{5R}(5qd_1 + d_2), \nonumber
\end{eqnarray}
where the functions $d_{i}$ are defined by the relations (we shall need $d_{3}$ slightly later) :
\begin{eqnarray}\label{d_functions}
d_1(q) & = & q (|q|-1)^2(|q|+2)\theta(|q|<1) , \nonumber \\
d_2(q) & = & (|q|-1)^2(-6|q|^3-12|q|^2+2|q|+1) \theta(|q|<1) , \\
d_3(q) & = & 35q|q|(|q|-1)^2(2|q|^2+4|q|-3)\theta(|q|<1) .  \nonumber
\end{eqnarray}
Parameter  $q = X_0/R$ in Eq.~(\ref{d_functions}) is the relative position of the center of mass of a harmonic. The $\theta$-function in Eq.~({\ref{d_functions}) is equal to one if its argument is a logical true and zero if it is a logical false. These functions arise because of the interatomic interactions between the two harmonics. Therefore, they are non-zero only when the harmonics are overlapping.

Using expectation values given by Eqs.~(\ref{Expect_values_XP}), (\ref{Expect_values_other}) and evaluating their dynamics with the help of Eq.~(\ref{Expectation_Equation}), results in a set of first order differential equations
describing the dynamics of the split condensates:
\begin{eqnarray}\label{Dless_Eqns_motion}
R_\tau &=& \frac{G}{R} ,                                                \nonumber \\
G_\tau &=& \frac{G^2}{R^2} - R^2\left(1-\frac{1}{2R^3}\right)+ \frac{1}{R} d_2(q), \nonumber \\
(X_0)_\tau &=& V + \frac{S}{10R}   ,                                    \\
\left(V + \frac{S}{10R}\right)_\tau &=& - X_0 + \frac{d_1(q)}{R^2}  ,  \nonumber     \\
S_\tau &=&  \frac{d_3(q)}{R} ,                                          \nonumber
\end{eqnarray}
where $A_\tau$ represents the derivative of the function $A$ with respect to time.

In deriving the equations of motion (Eq.~\ref{Dless_Eqns_motion}) , $\epsilon$  and  $V_{0}^{-1}$ were used as smallness parameters and terms of the order of $\epsilon^2$ and $V_0^{-2}$ have been neglected. This can be justified by the following estimate. For a BEC of $^{87} Rb$  atoms, $v_0 = 11.7 $ mm/s . For the longitudinal angular frequency $\omega = 2\pi \times 4.1 $ Hz ,  the angular frequency in the transverse dimensions $\omega_\perp = 2\pi \times 80 $ Hz  \cite{segal10}, and the number of atoms in the condensate $N = 10^4$ \cite{wang05},  the equilibrium size of a condensate $L_0$ given by  Eq.~(\ref{size}) is approximately $ 40 $ $\mu$m. For these parameters, the inverse of the dimensionless initial speed $V_{0}$ of the harmonics is $V_{0}^{-1} \approx 0.09$ and $\epsilon \approx  0.018$.

Finally, analysis of Eq.~(\ref{Dless_Eqns_motion}) requires evaluation of functions $D_{i}(q) = \int_{0}^{q}d_{i}(x)dx$, which are integrals of the functions $d_i(q)$, with respect to $q$:
\begin{eqnarray}\label{d_123_large}
D_{1}(q) & = &  \frac{1}{5}q^{2}(|q|^{3} - 5|q| + 5), \nonumber \\
D_{2}(q) & = &  - q(|q| - 1)^{3}(q^{2} + 3|q| + 1),   \\
D_{3}(q) & = &  \frac{1}{2}|q|^{3}(20q^{4} - 126 q^{2} + 175|q| - 70).\nonumber
\end{eqnarray}
Expressions for the functions $D_{i}(q)$ given by Eq.~(\ref{d_123_large}) are valid in the region $|q| < 1$. For $|q| \ge 1$, the functions $D_{i}(q)$ are constant and equal to their boundary values :  $D_{1}(\pm 1) = 1/5$, $D_{2}(\pm 1) = 0$, and $D_{3}(\pm 1) = - 1/2$.

\subsection{Evolution of the radius and the quadratic phase}
%
Time dependence of the radius $R(\tau)$ and the quadratic phase $G(\tau)$ can be obtained by solving the first two equations of Eq.~(\ref{Dless_Eqns_motion}). The contribution from the term with $d_{2}(q)$ in the equation for $G_{\tau}$
can be neglected and these equations reduce to :
\begin{eqnarray}\label{R_tautau}
R_{\tau \tau} = - R + \frac{1}{2R^2},
\end{eqnarray}
Integrating the Eq.~(\ref{R_tautau}) with initial conditions $R(0) = 1$ and $R_\tau(0) = 0$ yields
\begin{eqnarray} \label{R_tau}
R_\tau = \pm \sqrt{\frac{(1-R)(R^2+R-1)}{R}}.
\end{eqnarray}
The first of Eq.~(\ref{Dless_Eqns_motion}) then gives

\begin{eqnarray}\label{G_tau}
G(\tau) = \pm \sqrt{R(1-R)(R^2+R-1)}.
\end{eqnarray}

The analytic solution to  Eq.~(\ref{R_tau}) can be obtained in terms of elliptic functions.
It is important to notice that Eq.~(\ref{R_tau}) for $R$ (and, thus, Eq.~(\ref{G_tau}) for $G$) is ``universal'', i.e., independent of the trap frequencies, number of atoms in the condensate, etc., and needs to be solved only once.
Fig.~\ref{fig:G_and_R} shows the time evolution of $R $ and $G$ for a full trap period obtained by solving the first two of the Eq.~(\ref{Dless_Eqns_motion}) numerically.  The small kinks in the plot of $G$ during splitting, at recombination , and when the harmonics pass through each other, are due to mutual interaction between the two harmonics at overlap (the term with $d_{2}$ in equation for $G_{\tau}$).  Fig.~\ref{fig:G_and_R} shows that the neglect of this term in obtaining Eq.~(\ref{R_tau}) is an excellent approximation. It is interesting to note that the period of the oscillations of the radius is about $60\%$ of the trap period. The quadratic phase $G$ has the same period as $R$.
In our analysis we shall need the values of $R$ and $G$ only at the time of recombination $\tau = 2\pi$:  $R(2\pi) \approx 0.81$ and $G(2\pi) \approx 0.27$.
\begin{figure}
\includegraphics[width = 8.6cm]{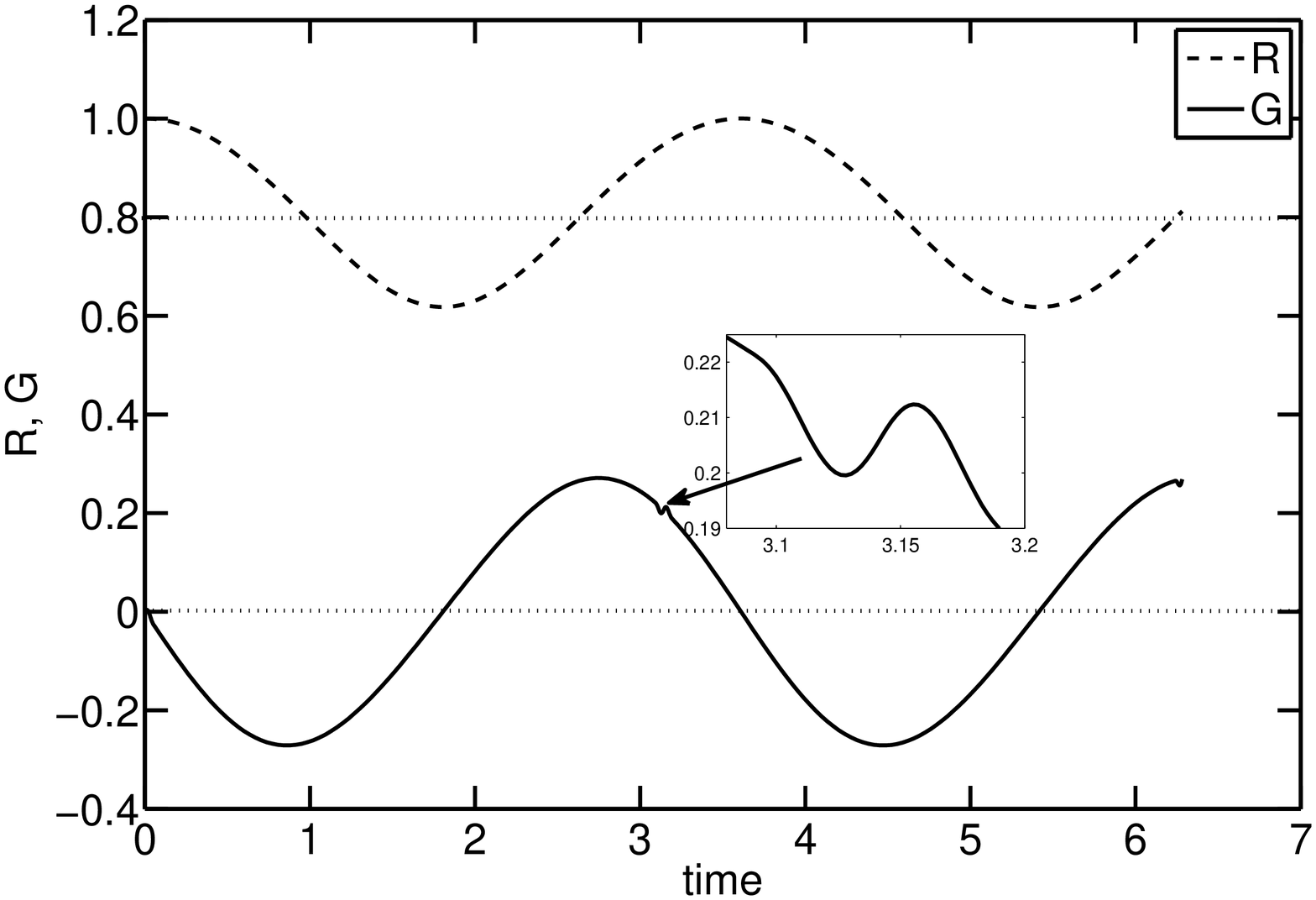}
\caption{\label{fig:G_and_R} Time evolution of the dimensionless radius $R$ of a harmonic  and the quadratic phase $G$ (rad) for a trap period. The horizontal axis is dimensionless time from $0$ to $2 \pi$. The inset picture shows the effect of interatomic interactions on $G$ when the two harmonics pass through each other.}
\end{figure}

\subsection{Evolution of the cubic phase}
Evolution of the cubic phase $S(q)$ is governed by the last of Eq.~(\ref{Dless_Eqns_motion}). The cubic phase changes only when the harmonics overlap because, otherwise, the function $d_3(q)$ in (Eq.~\ref{d_functions}) is zero. Integration of Eq.~(\ref{Dless_Eqns_motion}) yields
\begin{equation} \label{cubic_phase}
    S(q) = \int_{0}^{\tau} d\tau^{\prime} \frac{d_{3}(q)}{R} = \int_{0}^{q} dq \frac{d_{3}(q)}{R}\frac{d\tau}{dq} \approx  \frac{D_{3}(q)}{V_0},
\end{equation}
because in the region of overlap $ d q/ d\tau \approx \pm V_{0}/ R$  and $S(0) = 0$.
The function $D_3(q)$ is given by Eq.~(\ref{d_123_large}). After the first separation of the harmonics, the value of $S$ outside the overlap region is
\begin{equation}\label{S0_1}
    S (1) = - \frac{1}{2V_{0}},
\end{equation}
because $D_3(1) = -1/2$ . The difference between the values of $S$ before and after the passage of the harmonics through each other around mid-cycle $\tau \approx T/2$ is zero , since its calculation involves integration of the odd function $d_{3}(q)$ from $q = 1$ to $q = -1$.
Finally, near the end of the cycle
\begin{equation}\label{S0_2}
    S(q) - S(-1) = \frac{1}{V_{0}}\left[D_{3}(q) - D_{3}(-1)\right].
\end{equation}
Combining Eqs.~(\ref{S0_1}) and (\ref{S0_2}), the value of $S$ near the end of the cycle
\begin{equation}\label{S_end_norefl}
    S(\tau \approx 2\pi) = \frac{D_{3}(q)}{V_{0}} \approx - \frac{35 |q|^{3}}{V_{0}}.
\end{equation}
in the lowest order of $|q|$.

Fig.~\ref{fig:S} shows the evolution of $S$ with time. It is zero initially and grows to a negative peak once the two harmonics start moving away from each other. After the harmonics completely separate, the value of $S$ remains constant at its boundary value. The inset picture in Fig.~\ref{fig:S}  shows the evolution of $S$ when the two harmonics pass through each other.

\begin{figure}
\includegraphics[width = 8.6cm]{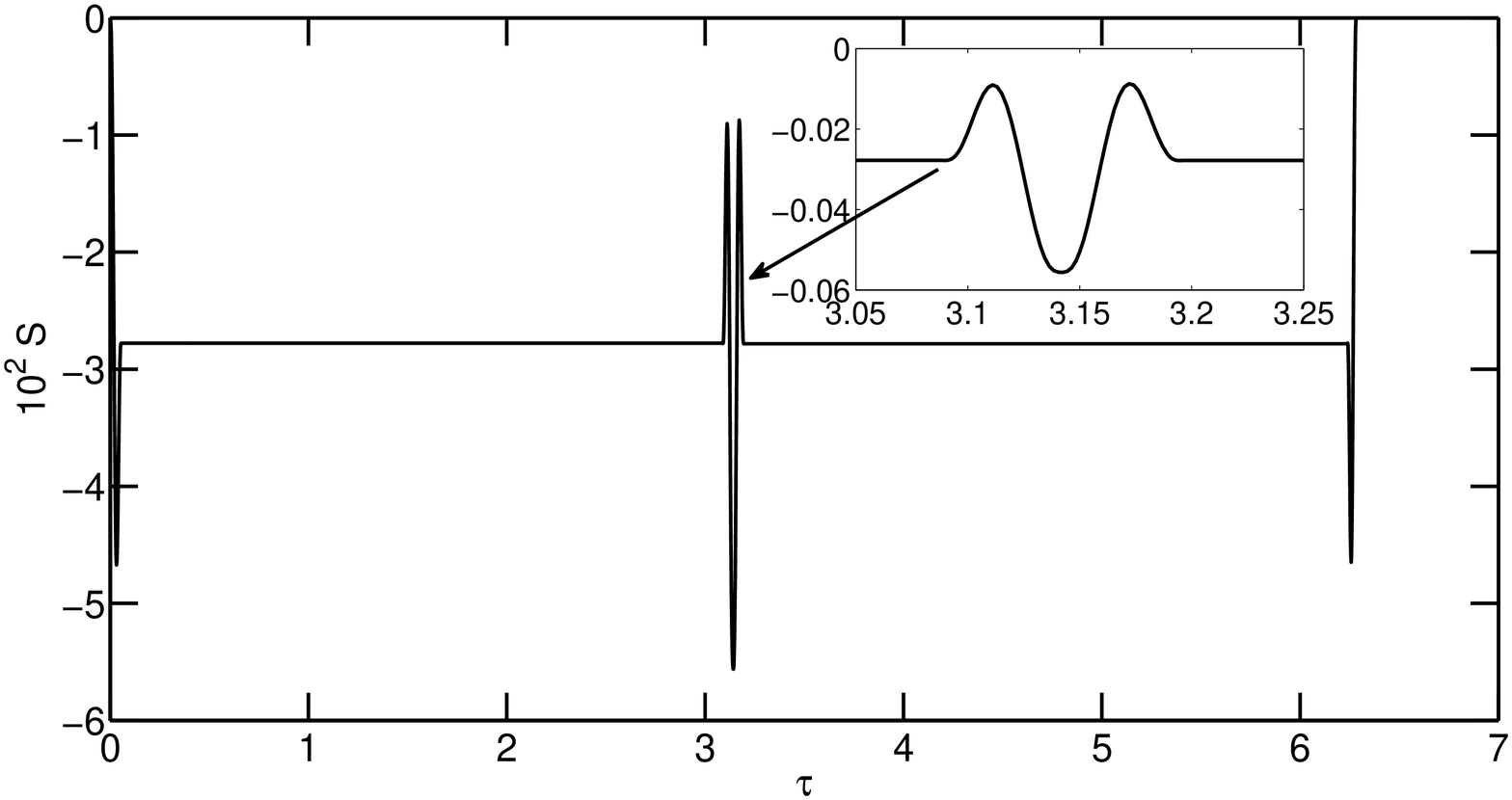}
\caption{\label{fig:S} Time evolution of the cubic phase $S$(rad). The horizontal axis is dimensionless time, $\tau$ from $0$ to $2 \pi$. The cubic phase develops only when the clouds overlap during splitting, when they pass through each other and when they recombine. The inset picture shows the evolution of $S$ when the clouds pass through each other. }
\end{figure}

\subsection{Energy of the system}

The total energy $E$ of the condensate is obtained by evaluating the expectation value of the Hamiltonian $\langle \psi | H| \psi \rangle $ with respect to the total wave function $\psi$ of the split condensates (Eq.~\ref{split_wavefn}). This quantity can be represented as the sum of the three terms:
\begin{eqnarray}\label{total_energy}
E = E_K + E_P + E_N ,
\end{eqnarray}
with
\begin{eqnarray}
E_K &=& \frac{1}{2} \left(V+ \frac{S}{10R}\right)^2 + \frac{G^2}{10R^2},\nonumber \\
E_P &=& \frac{1}{2}\left(X_0^2+\frac{R^2}{5}\right), \nonumber \\
E_N &=& \frac{\left(1 + 10qd_1+2d_2\right)}{10R},\nonumber
\end{eqnarray}
where the $d$ - functions are given by the Eq.~(\ref{d_functions}). The term $E_K$ in Eq.~(\ref{total_energy}) is the kinetic energy of the system. It depends upon the speed of the clouds, the quadratic and the cubic phases. The kinetic energy right after splitting is proportional to $V_0^2$ but, when the clouds evolve with time, the quadratic and the cubic phases develop and the kinetic energy has terms containing speed as well as these phases.  The term  $E_P$ is the potential energy of the clouds. This energy is equal to the sum of the potential energy of the condensate in the trap caused by its finite size,  and the potential energy due to the displacement of the center of mass of each cloud after splitting. Due to its finite size, a cloud has non-zero potential energy even when it is at the bottom of the trap. Finally, the term $E_N$ is nonlinear energy due to atom-atom interactions in the condensate.

During the interferometric time between the splitting and recombination pulse, the total energy of the system is conserved.

\subsection{Wave function at recombination}
The harmonics are recombined by using a recombination pulse at the end of the interferometric cycle time,  $T$. Since we are considering the case of $(\phi_0)_\pm = 0$ in Eq.~(\ref{density_phase}), the recombination pulse, in the ideal situation, should recombine the two harmonics into one at rest. But, because of the spatially-dependent phases accumulated during the interferometric cycle, there will be three harmonics in the output ports - one at rest and two moving in opposite directions \cite{stickney07}. The wave function for the zeroth order harmonic at recombination is given by the expression
\begin{eqnarray}\label{Recom_wave_function}
\psi_0(\xi,q) = \sqrt{n(\xi)}\cos\phi(\xi,q),
\end{eqnarray}
where the spatial relative phase across a harmonic is
\begin{eqnarray} \label{spatial_phase}
\phi (\xi,q)  = \Delta K(q) \xi + \Gamma(q) \xi^{3}.
\end{eqnarray}
The strengths of the linear and cubic phases as functions of $q$ is given by the relations
\begin{eqnarray}\label{DeltaK_Gamma}
\Delta K(q) & = & \frac{R_T}{\epsilon}\left(V_T-V_0 - \frac{G_T}{R_T}q + \frac{S_T}{2R_T}q^2\right), \nonumber \\
\Gamma(q) & = &  \frac{S_T}{6\epsilon},
\end{eqnarray}
where $R_T$, $V_T$ , $G_T$ and $S_T$ are evaluated at time $T$. In  Eqs.~(\ref{Recom_wave_function}) and ~(\ref{spatial_phase}), $\xi = X/R_T$ is the normalized coordinate and $q = X_0/R_T$ is the normalized position of the center of mass. We have neglected a small degree of incomplete overlap in the densities,  but have taken it into account in phases.
%
\section{Fringe contrast and limits of performance}\label{sec:limit}

The population in the zeroth order harmonic $N_{0}$ is given by the expression
\begin{equation}\label{Relative_population}
    N_{0} = N_{tot}\int |\psi_{0}(\xi,q)|^{2} d\xi,
\end{equation}
where $\psi_{0}$ is defined by Eq.~(\ref{Recom_wave_function}). In the ideal case $\Delta K = \Gamma = 0$, the contrast of the interference fringes $V = 2N_{0}/N_{tot} - 1$ should be equal to one,
but the coordinate-dependent phase results in a decrease of the contrast. The physical reason for this decrease is explained by Fig. \ref{fig:LargeDeltaK}. When the coordinate-dependent phase $\phi(\xi,q)$
becomes large, the density profile undergoes several oscillations between its maximum value (given by the envelope $n(\xi)$) and zero.  In this limit, the relative population in the zero momentum state
after recombination $N_0/N_{tot} = \int d\xi |\psi_0(\xi)|^2$ approaches $1/2$, and the contrast $V = [2(N_0/N_{tot}) - 1]$ goes to zero.

The contrast of the fringes for small values of $\Delta K$ and $\Gamma$ can be expressed as
\begin{eqnarray}\label{contrast}
V  = 1 - \frac{2}{5}\left[(\Delta K)^{2} + \frac{6}{7}\Delta K \Gamma\ + \frac{5}{21}\Gamma^{2}\right].
\end{eqnarray}
Stickney et al. \cite{stickney08} noticed that the best contrast
does not necessarily correspond to the complete overlap of the BEC clouds $q = 0$ and sometimes can be improved by slightly changing the recombination time (or, equivalently, the value of $q$).
We shall evaluate the limits of performance of the interferometer by minimizing the bracketed quantity on the right hand side of Eq.~(\ref{contrast}) with respect to $q$ and setting the result (somewhat arbitrarily) to $0.5$.

The velocity $V_{T}$ of the cloud at recombination time in the expression for $\Delta K$ given by Eq.~(\ref{DeltaK_Gamma}) can be obtained by equating the total energy of the clouds at recombination time $ T$ to their total energy right after splitting ($\tau =0$) because the total energy of the system is constant. From  Eq.~(\ref{total_energy}), the total energy of the system at time $\tau = 0$
\begin{eqnarray}\label{energy_t0}
E (0) = \frac{1}{2} \left(V_0^2+\frac{4}{5}\right),
\end{eqnarray}
and the total energy at recombination time, $T$
\begin{eqnarray} \label{energy_tT}
E (T) = \frac{V_T^2}{2}  + \frac{G_T^2}{10R_T^2}+\frac{R_T^2}{10}+ \frac{3}{10R_T} .
\end{eqnarray}
Equating the right hand sides of  Eqs.~(\ref{energy_t0}) and (\ref{energy_tT}) gives
\begin{eqnarray}\label{DeltaV}
V_T-V_0 \approx  - \frac{1}{2V_0}\left(\frac{G_T^2}{5R_T^2}+\frac{R_T^2}{5}+\frac{3}{5R_T}- \frac{4}{5}\right),
\end{eqnarray}
where $V_T$ is the velocity of the harmonic right before the recombination pulse. By substituting $(V_T-V_0)$ from  Eq.~(\ref{DeltaV}) into the first equation of  Eq.~(\ref{DeltaK_Gamma}), one gets
\begin{eqnarray}\label{DeltaK_final}
\Delta K (q) = - \frac{1}{\epsilon} \left(\frac{0.04}{V_0} + 0.27 q\right),
\end{eqnarray}
where $q = X_0/R_T$ is the relative position of the center of mass of the $\psi_+$ harmonic. To get $\Delta K$ in the form given by  Eq.~(\ref{DeltaK_final}), we used $R_0 = 1$, $R_T = 0.81$, and $G_T \approx 0.27 $.
Similarly, from the second equation of Eq.~(\ref{DeltaK_Gamma}) and Eq.~(\ref{S_end_norefl}),
\begin{eqnarray}\label{Gamma}
\Gamma (q) = - \frac{35}{6\epsilon V_0}|q|^3 .
\end{eqnarray}
Minimizing the bracketed quantity in the right hand side of the Eq.~(\ref{contrast}) with respect to $q$ and requiring $V \ge 1/2$ results in an inequality
\begin{equation}\label{good_op_norefl_1}
    \frac{0.01}{\epsilon V_{0}^{4}}  \le 1
\end{equation}
 that gives a working region in the parameter space of the interferometer. The inequality (\ref{good_op_norefl_1}) can be expressed in terms of the dimensional experimental parameters as follows:
\begin{eqnarray}\label{working_region}
    \left(\frac{\hbar \omega_\perp^2 \omega a_s^2}{10 Mv_0^4}\right)^{1/2}N \le 1,
\end{eqnarray}
where $N$ is the number of atoms in the trap of axial angular frequency $\omega$ and the transverse frequency $\omega_\perp$. This inequality gives a fundamental limit on performance of a guided BEC-based free oscillation interferometer in Michelson-type geometry. The second fundamental limit is due to phase diffusion, which can not be described in the mean field approach. In addition, there can be technical limitations like the noise (caused by vibrations), misalignment of the splitting laser pulse, etc.

Figure.~\ref{fig:Working_Region} shows the working region of a free oscillation interferometer for a transverse trapping frequency $\omega_\perp = 2\pi \times 80 \ Hz$. In the region below the boundary line (which has been obtained taking the equality sign in Eq.~(\ref{working_region})) the interferometric contrast exceeds $50\%$. The maximum number of atoms corresponding to the boundary region for a given trap can be read directly from the graph. For example,  for $\omega = 2 \pi \times 4.1 \ Hz$, $N \approx 10^6$ and the interferometric cycle time is $244 \ ms$ (the trap period).

\begin{figure}
\includegraphics[width=8.6cm]{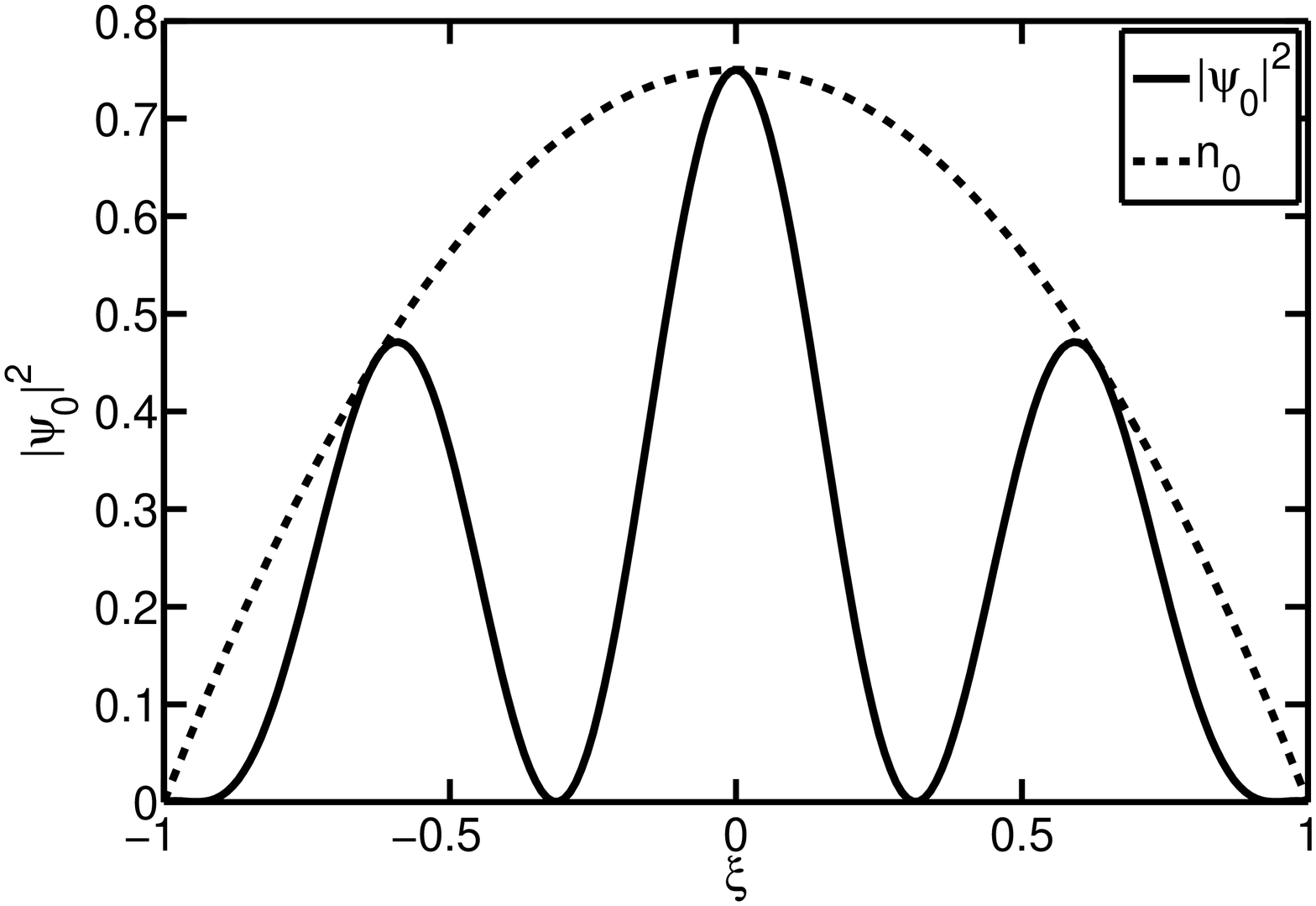}%
\caption{\label{fig:LargeDeltaK}The probability density $|\psi_0|^2$ versus the coordinate $\xi$ for $\Delta K = 5$ and $\Gamma = 0$. The probability density oscillates several times under its envelope that reduces the contrast of the interference fringes.}
\end{figure}

\begin{figure}
\includegraphics[width=8.6cm]{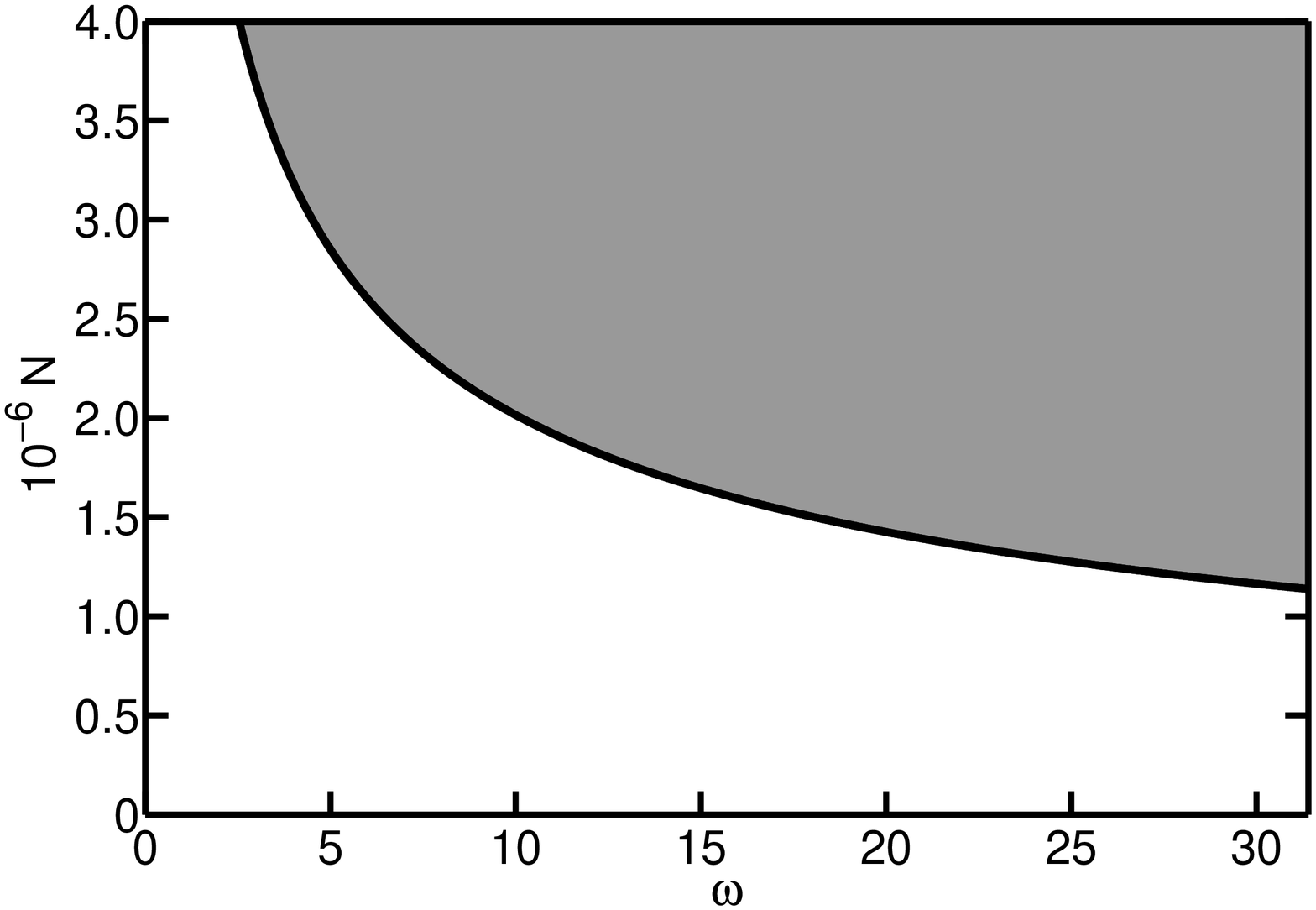}%
\caption{\label{fig:Working_Region}Working region in parameter space of a free oscillation interferometer , with the longitudinal trap frequency $\omega$ (rad/s) and the total number of atoms $N$ in the condensate. The interferometer works in the unshaded region and does not work in the shaded region.}
\end{figure}

\section{Discussion of single - and double reflection geometries}\label{sec:single-double}

In this section, we briefly re-derive results of the analysis of the single - and double reflection atom Michelson interferometers \cite{stickney07, stickney08} and compare their performance with that of the free oscillation interferometer. In a single reflection atom Michelson interferometer, a BEC sitting at the bottom of a weakly-confining harmonic trap is split into two harmonics, which move in opposite directions with the velocities $\pm V_0$. At time $\tau = T/2$ where $T$ is the interferometric cycle time,  a reflection pulse is applied, which adds velocities of $\mp 2V_0$ to the harmonics so that the $\psi_+$ harmonic now moves with a velocity $(V_+ - 2V_0)$ and the $\psi_-$ harmonic moves with a velocity $(V_- + 2V_0)$, where $V_\pm = \pm V$  are the velocities of the two harmonics right before the first reflection pulse is applied. A recombination pulse, identical to the splitting pulse is used to recombine the two harmonics at time $\tau = T$.

The total energy of the system is given by  Eq.~(\ref{total_energy}). Unlike in the case of a free oscillation interferometer, the energy is not conserved for a complete cycle because of the reflection pulses , but remains constant between them.  In this case, the velocity of the harmonics at the time of recombination can be calculated from Eq.~(\ref{total_energy}) taking into account the momentum kick given to the harmonics by the reflection pulse.
The difference between the final speed of the $\psi_+$ harmonic and its initial speed $\Delta V = (V_T - V_0)$ in the dimensionless variables turns out to be:
\begin{eqnarray}\label{DeltaV_single}
\Delta V  \approx \frac{V_0T^2}{4}.
\end{eqnarray}
In deriving Eq.~(\ref{DeltaV_single}), Eqs.~(\ref{R_tau}) and (\ref{G_tau}) have been solved to the lowest order in $\tau$ yielding $G \approx - \tau/2$ and $R\approx 1 - \tau^2/4$, because the duration of the interferometric cycle in this case is much less than the full period of oscillation in the trap. The dimensional version of Eq.~(\ref{DeltaV_single}) reads:
\begin{eqnarray}\label{DeltaV_single_final}
\Delta v_D  \approx \frac{v_0}{4}(\omega T_D)^2,
\end{eqnarray}
where $v_0 = 2\hbar k_l/M$, $L_0$ is the equilibrium size of the condensate and $T_D$ is the dimensional time for the interferometric cycle.  Relation (\ref{DeltaV_single_final}) coincides with the analogous relation (Eq. 24) in  Ref.~\cite{stickney08} obtained by a different technique.

In a double reflection interferometer, two reflection pulses are applied at times  $\tau=T/4$ and $\tau=3T/4$ before the recombination pulse is applied at the end of the interferometric cycle time $\tau = T$.
Using the procedure analogous to that described above, we calculate the difference between the velocity of the $\psi_+$ harmonic at recombination $(V_T)$ and initial velocity $(V_0)$ to be equal to
\begin{eqnarray}\label{DeltaV_double}
\Delta V \approx - \frac{3T^2}{20V_0},
\end{eqnarray}
or, in dimensional variables,
\begin{eqnarray} \label{DeltaV_newpaper}
\Delta v_D \approx  - \frac{3}{20} \frac{\omega^4T_D^2L_0^2}{v_0}.
\end{eqnarray}
This expression matches with the analogous relation (Eq. 43) in Ref.~\cite{stickney08}.

Equations (\ref{DeltaV_single}) and (\ref{DeltaV_double}) have been obtained in the limit when the clouds have zero spatial overlap at the time of application of the reflection pulses.

To compare the performances of the single-, double-reflection, and free oscillation interferometers, it is enough to compare the differences between the velocity of the $\psi_+$ harmonic at recombination $(V_T)$ and its initial velocity $(V_0)$ in the three geometries. For a free oscillation interferometer, $(V_T - V_0)$ given by the Eq.~(\ref{DeltaV}) becomes
\begin{eqnarray} \label{DeltaV_free_dimensional}
\Delta v_D \approx  - \frac{1}{20} \frac{\omega^2 L_0^2}{v_0}
\end{eqnarray}
in dimensional variables.
A comparison of the Eqs.~(\ref{DeltaV_single_final}), ~(\ref{DeltaV_newpaper}) and ~(\ref{DeltaV_free_dimensional}) shows that the difference in velocities at recombination and initial velocity of a cloud is much larger in a single reflection interferometer, less in double reflection interferometer and much smaller in a free oscillation interferometer for a given trap frequency. Since this velocity difference is the main cause of the loss of contrast (cf.  Eqs.~(\ref{DeltaK_Gamma}) and (\ref{contrast})), an increasingly improved contrast can be obtained in a double reflection and free oscillation interferometers compared to a single reflection interferometer.

\section{Conclusions}\label{sec:conclusions}
In this paper, we analyzed the operation of a BEC-based free oscillation interferometer with optical splitting and recombination of the BEC clouds.
Our one-dimensional (1D) analytical model is based on the mean field approximation in the Thomas-fermi limit. From the 1D
Gross-Pitaevskii equation, we derive a closed set of ordinary differential equations for the parameters describing the shape
of the density envelope and the spatially-varying phase of the BEC clouds. The derivation is based on the
equations of motion for the quantum-mechanical expectation values associated with these parameters. The main result of the paper is Eq.~(\ref{working_region}), which gives the working region of the interferometer in the parameter space and shows how the performance of the interferometer depends on different parameters of the experiment such as the number of particles, longitudinal and transverse frequencies of the trap, and the velocity imparted by the splitting laser pulses. According to our analysis, the reason for the loss of the coherence in a free oscillation interferometer is oscillations of the density envelopes of the clouds with a period different from the longitudinal period of the trap.

The analysis of the paper does not include effects beyond the mean field approximation such as finite-temperature phase fluctuations along the length of the elongated BEC clouds and phase diffusion. Ref.~\cite{stickney08} discussed the importance of the phase fluctuations and concluded that they are negligible for the parameters of the recent experiments \cite{wang05, garcia06, burke08, segal10}. The phase diffusion, specifically in the context of atom interferometers with the optical splitting and recombination of the clouds, has been recently analyzed in \cite{ilo-okeke10}. Results of this analysis, applied to the case of a free oscillation interferometer, predict that the region of good performance is given by the inequality
\begin{equation}\label{phase_diffusion}
    \left(\frac{a_{s}}{\bar{a}}\right)^{2/5}\left(\frac{2\pi \bar{\omega}}{\omega}\right)N^{-1/10} \le 1,
\end{equation}
where $\bar{\omega} = (\omega_{\perp}^{2}\omega)^{1/3}$ and $\bar{a} = \sqrt{\hbar/(M\bar{\omega})}$.

The model of Ref.~\cite{ilo-okeke10} goes beyond the mean field approximation by accounting for the mode-entangled nature
of the two BEC clouds after the splitting, but does not account for the development of spatially-varying phases caused by atom-atom interaction during the propagation, as opposed to the present paper. Thus, the physics behind Eqs.~(\ref{working_region}) and (\ref{phase_diffusion}) is complementary, and both these inequalities have to be evaluated and their values compared for any particular experiment.

The relative importance of the effects due to spatially-varying phases caused by atom-atom interactions and the phase diffusion is given by the left-hand sides of Eqs.~(\ref{working_region}) and (\ref{phase_diffusion}), respectively.  The left-hand-side of Eq.~(\ref{working_region}) for the parameters of the experiments by Burke et al. \cite{burke08} and Horikoshi et al.~\cite{horikoshi07} is much less than one, and equals about $0.8$ in the experiments by Segal et al.~\cite{segal10}. The left-hand-side of Eq.~(\ref{phase_diffusion}) for Ref.~\cite{burke08} is small as compared to one and equals to about $0.65$ and $1.0$ for Refs.~\cite{horikoshi07} and \cite{segal10}, respectively. This shows that the phase diffusion could be partially responsible for the degradation of coherence in \cite{horikoshi07} and that both the effects discussed in this paper and the phase diffusion could be at least partially responsible for the loss of contrast in the experiments \cite{segal10}. The authors of Refs.~\cite{horikoshi07,burke08,segal10} also list vibrations as a cause for the degradation of the coherence.

In the experiments discussed in the paper, the frequency of the trap along the guiding direction is much less than those along the transverse directions. The BEC clouds are cigar-shaped with the largest dimension along the weak guiding direction of the trap and are moving along the same direction. This is the reason why a 1D theory is a good approximation to the experimental situation. A possible slight misalignment of the optical splitting pulses can result in a more complicated 2D or 3D motion of the BEC clouds and their rotations. Analysis of such dynamics requires generalization of the 1D model of the present paper to higher dimensions.


\section{Acknowledgments}
This work was partially supported by the Defense Advanced Research Projects Agency (Grant No. W911NF-04-1-0043).


\bibliographystyle{apsrev}


\end{document}